\documentclass[%
reprint,
 amsmath,amssymb,
pra,
floatfix,
]{revtex4-1}

\usepackage{graphicx}
\usepackage{float}
\usepackage[caption=false]{subfig}
\usepackage{dcolumn}
\usepackage{bm}
\usepackage{hyperref}
\usepackage[mathlines]{lineno}
\usepackage{import}
\usepackage{longtable}
\usepackage[table]{xcolor}
\usepackage{array}
\newcolumntype{?}{!{\vrule width 1pt}} 

\begin{document}

\preprint{APS/123-QED}

\title{Fluid and Thermal Dynamical Analysis of Three-Dimensional Spin-Exchange Optical Pumping Cells}

\author{G.M. Schrank}




\date{May 17, 2020}

\begin{abstract}
We present a finite-element analysis of the fluid dynamics, thermal dynamics, and alkali diffusion in a common cell geometry used for spin-exchange optical pumping of $^{129}$Xe using a flow-through polarizer design. The analysis is the first to simulate aspects of the laser run-away effect observed in some optical pumping cells. The analysis further suggests that high-temperature gas in the outlet tube may give rise to rapid depolarization due to high concentrations of alkali vapor. Finally, the analysis indicates that the alkali number density distribution and the specifics of the dynamics are highly dependent upon the the distribution of the alkali metal in the optical pumping portion and outlet tube of the cell. 
\end{abstract}

\pacs{Valid PACS appear here}
\maketitle


\section{\label{sec:intro}Introduction}
Spin-exchange optical pumping (SEOP) is a technique which allows for  the nuclear spin-angular momentum of certain noble gasses to be increased to of order 10\%. SEOP has been used to study the NMR characteristics of porous media \cite{Terskikh2002AMaterials}, to examine protein dynamics \cite{Schroder2013XenonAlert}, and to perform clinical lung-imaging in humans \cite{Oros2004HyperpolarizedMRI}

 SEOP involves two steps: optical pumping of an alkali metal vapor and spin-exchange from the alkali metal vapor to the noble gas nuclei. First, a beam of circularly polarized light is directed on to a transparent cell containing a macroscopic amount of alkali metal, usually rubidium (Rb). The cell is heated to between 100-200 $^{\circ}$C to produce an optically-thick alkali-vapor. The laser is tuned to the D1 transition frequency, and the circular polarization of the laser imposes a selection-rule that will only excite electrons from one spin-state to the other. The spin-polarized electron quickly recouples with the nuclear spin, and thus contributes to the spin-polarization of the atom.  

Next, during spin-exchange, the alkali vapor transfers spin-angular momentum to the noble gas nuclei. A gas mixture is introduced into the cell containing a noble gas and some other inert gases, usually helium and nitrogen. During collisions, the alkali metal electron couples with the noble gas nuclei via the Fermi-contact interaction, thus transferring its spin-angular momentum. Although the alkali metal atom leaves this interaction depolarized, it is quickly repolarized by absorption of another photon from the laser.  

Hyperpolarization of xenon-129 ($^{129}$Xe) using Rb vapor has become a popular method of producing polarized gas, both because $^{129}$Xe is inexpensive relative to other options and because SEOP of $^{129}$Xe can be accomplished on a relatively fast time-scale. Hyperpolarized xenon-129 (HP $^{129}$Xe) gas is typically produced in a continuous manner using a flow-through polarizer \cite{Driehuys1996High-volume129Xe, Ruset2006Optical129Xe, Schrank2009a}. A flow-through polarizer operates by flowing a $^{129}$Xe gas mixture through an optical pumping cell containing Rb vapor. Because $^{129}$Xe can undergo spin-exchange on a fast-time scale, the gas can flow through cells with total gas flow rates $\sim$1 SLM. $^{129}$Xe is usually kept at concentrations between 1-5\% in these systems, thus a flow-through polarizer can reasonably produce 1 liter of HP$^{129}$Xe in approximately an hour.  

With increased interest in this technique for medical imaging, there is a need to develop technology which can reliably produce high-volume, high-polarization HP$^{129}$Xe on short timescales. Recently, Freeman, et al. \cite{Freeman2014} noted that some styles of flow-through polarizers were not producing HP $^{129}$Xe with the polarization that theoretical models predicted. In order to gain insight into this deficiency, a new Finite Element Model (FEM) model, which builds upon previous models \cite{Fink2005, Fink2007}, was constructed to simulate the full three-dimensional dynamics of SEOP cells similar to those examined by Freeman \cite{Schrank2019ACode}. Here, we present the preliminary results of that model.
\section{Setup of the Simulation\label{sec:simsetup}}
\begin{figure*}
    \subfloat[\label{fig:fullcell}]{\includegraphics[width=0.4\textwidth]{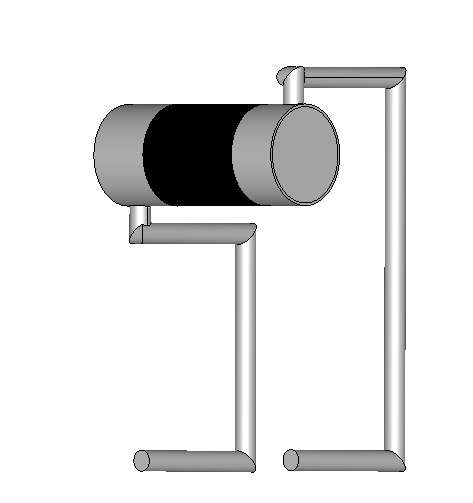}}
    \subfloat[\label{fig:fullcellvert}]{\includegraphics[width=0.4\textwidth]{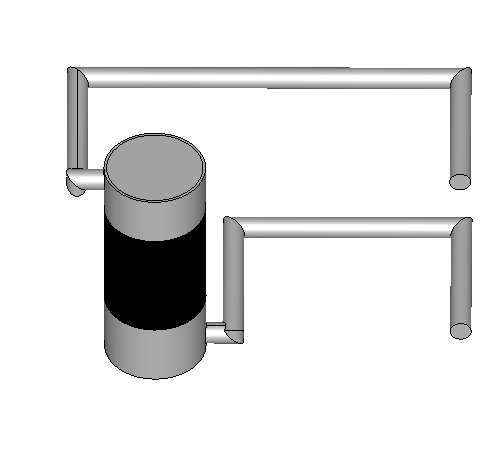}}\\
    \subfloat[\label{fig:halfcell}]{\includegraphics[width=0.4\textwidth]{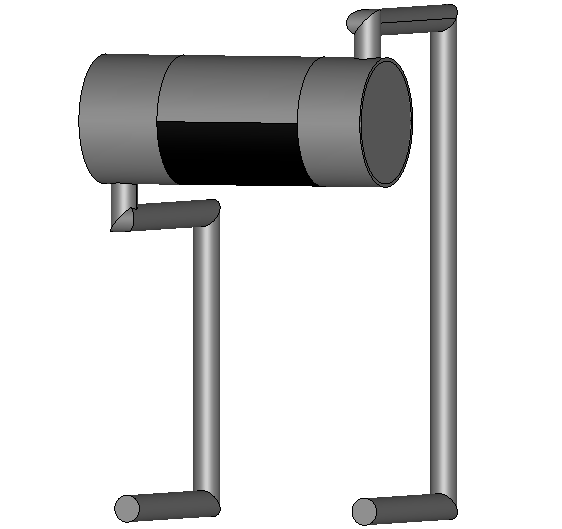}}
    \subfloat[\label{fig:rbdropcell}]{\includegraphics[width=0.4\textwidth]{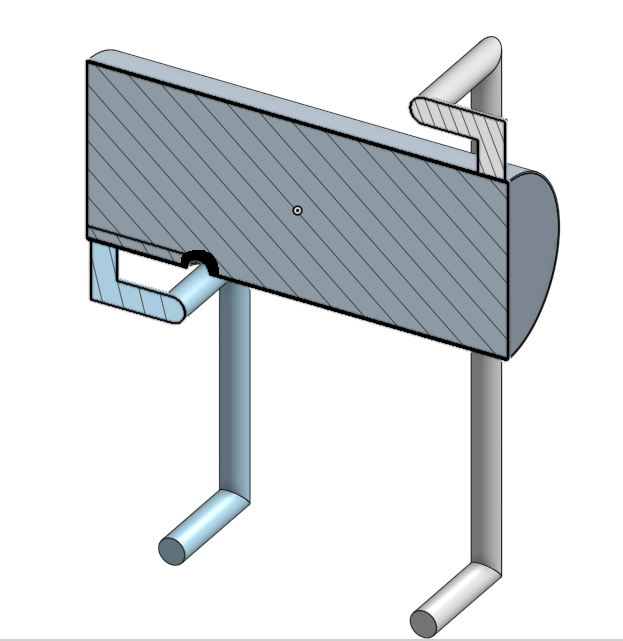}}
    \caption{The different geometries that were modelled in the current study: (\ref{fig:fullcell}) the 100-cc horizontal geometry with a full-circumferential shell for the Rb source, (\ref{fig:fullcellvert}) the 100-cc vertical geometry with a full-circumferential shell for the Rb source, (\ref{fig:halfcell}) the 100-cc horizontal geometry with a half-circumferential shell for the Rb source, (\ref{fig:rbdropcell}) the 300-cc horizontal geometry with a drop for the Rb source. The 300-cc cell has an area drawn directly behind the Rb drop to model the drops shadow. A further 100-cc vertical geometry with a half-circumferential shell for the Rb source was modelled but not pictured in the figure. In the figures, areas in black represent the Rb sources in the optical pumping section of the cells. The outlet tube sources are not pictured in the figures.}
    \label{fig:allgeometries}
\end{figure*}
Three different geometries where investigated in the course of this study: (1) a 100-cc cell with a full-circumferential Rb distribution, (2) a 100-cc cell with a half-circumferential Rb distribution, and (3) a 300-cc cell with a Rb drop (see Figure \ref{fig:allgeometries}). The geometries were designed to closely resemble the geometries of experimental cells used by Freeman et al. However, it should be noted that Freeman et al. did not report the Rb distribution in the cells, and this may strongly alter the simulated results.

Rb sources for the diffusion module were modelled in two areas in the optical pumping cell geometries. First, a Rb source was modelled in the optical pumping body (cylindrical portion). This source was either modelled as a thin film (Figure \ref{fig:fullcell}, \ref{fig:fullcellvert}, and \ref{fig:halfcell}) or a drop (Figure \ref{fig:rbdropcell}), and these sources were meant to emulate the Rb metal that was introduced into the body of the optical pumping cells for Freeman et al.'s testing. 

The 300-cc cell's Rb-drop source proved to be a special challenge for the model. The FEM model for the laser absorption is very poor at handling discontinuities, and a discontinuity exists directly behind the Rb drop: a shadow. The Rb drop was modelled to have a 3.5-mm radius, and thus it extended up above the wall of the cell. In order to account for this shadow cast by the drop, a second region behind the drop had to be drawn where the laser absorption model was disabled. This effectively caused that area to be in the dark. Models that did not employ a shadow in the geometry would not execute.

The second source/sink was modelled as a thin film in the outlet tube of the cell. In the 100-cc cells, this thin film started at the joint between the outlet tube and the cell body, and it extended to the final bend in the outlet tube. Two different configurations were used for the Rb source in the outlet tube of the 300-cc cell: (1) the same configuration as the 100-cc cells, and (2) the thin film started at the first bend in the outlet tube and extended to the final bend in the outlet tube. This source/sink was meant to emulate the area of the cell where Rb vapor condensed after exiting the optical pumping region. Visual inspection of used optical pumping cells of this design show buildup of Rb metal in the outlet tube of the cell.  

A body force, that is a condition applied to the entire simulated volume as opposed to just at a boundary, was applied to the models in order to simulate gravity and enable gas convection in the simulation. This force was modelled either such that the cell was oriented horizontally (Figure \ref{fig:fullcell}) or vertically (Figure \ref{fig:fullcellvert}). The vertical case was only investigated in the 100-cc cells.

Current flow-through polarizer designs that use cells of this size exclusively use a horizontal configuration. Therefore, the vertical simulations should be viewed as a more exploratory simulation rather than an attempt to understand current cell dynamics. However, some flow-through polarizers with much larger cells than those explored in this study do use a vertical orientation

Cells were modelled at four wall-temperature boundary conditions: \hbox{110 $^{\circ}$C}, \hbox{120 $^{\circ}$C}, \hbox{130 $^{\circ}$C}, and \hbox{140 $^{\circ}$C}. These different boundary conditions simulated different oven-air temperatures that the optical pumping cell could experience. Other laser, cell, flow-rate, and HP$^{129}$Xe relaxation rate configuration parameters are identical to those described in Ref. \cite{Schrank2019ACode}, as are the parameters and expressions used for important physical properties such as the diffusion constant, etc. 
 
The computational parameters of the model and the model implementation are also described in Ref. \cite{Schrank2019ACode}. All the successful simulations were run as transient simulations with 600 time steps at 0.1 sec. per time step. The time to compute a single transient model was tens of days of CPU time. There were several attempts to run steady-state simulations on some of the 300 cc geometries. However, all of the attempts failed to produce converging results.

The module that calculates the $^{129}$Xe polarization was disabled for transient simulations of the 100 cc model because of an instability in the computational module that was observed in transient-model tests. The 300 cc model did not suffer from these computational difficulties, and thus the $^{129}$Xe polarization module was enabled during these calculations. Transient simulations were run until changes in the calculated (1) Rb polarization, (2) average cell-body temperature, and (3) laser absorption were less that 0.05\%. If those conditions did not occur, the simulations were run until either the simulation completed the programmed number of time-steps (600; corresponding to 60 seconds of simulated time) or the simulation terminated with a failure to converge. 

Transient simulations that reached a steady state in all three metrics were used as the initial conditions in a steady state simulation with the $^{129}$Xe polarization module enabled in order to calculate the output polarization of HP$^{129}$Xe. The output polarization of HP$^{129}$Xe was calculated by averaging the polarization in the cell outlet tube over a slice 4 cm from the edge of the geometry. The slice was offset from the edge of the geometry because of artifacts in the simulations caused by the boundary.

Some of the solutions reached a computational point where the simulation would not restart, presumably because the solution previous time-step's solution vector was poorly conditioned.
\section{Results\label{sec:results}}
\begin{table*}[t]
\caption{Results of the 100-cc simulations at all temperatures. The top row describes the four different Rb configuration and cell orientation combinations used in the simulations. In the case that the simulation reached steady state, the predicted HP$^{129}$Xe polarization and laser absorption are reported. For simulations that did not reach steady state, the results are described in three ways: (1) full oscillations were observed, (2) initial oscillations were observed, (3) a rapid rise in temperature and laser absorption were observed. Text in italics denotes premature termination of the simulation.\label{tab:all100ccresults}}
\begin{tabular}{c|c|c?c|c|}
    &\textbf{Full-Horz.}    &\textbf{Full-Vert.}    &\textbf{Half-Horz.}         & \textbf{Half-Vert.}\\ \hline
\textbf{110 $^{\circ}$C}&\begin{tabular}{c c}$^{129}$Xe Polarization:&4.7\%\\Laser Absorp.:&7.9\%\end{tabular}&\begin{tabular}{c c}$^{129}$Xe Polarization:&5.5\%\\Laser Absorp.:&2.2\%\end{tabular}&\begin{tabular}{c c}$^{129}$Xe Polarization:&2.0\%\\Laser Absorp.:&4.5\%\end{tabular}&\begin{tabular}{c c}$^{129}$Xe Polarization:&2.3\%\\Laser Absorp.:&4.5\%\end{tabular}\\ \hline
\textbf{120 $^{\circ}$C}&Oscillations&Oscillations& \textit{Rapid Temp. and Abs. Increase} &Oscillations\\ \hline
\textbf{130 $^{\circ}$C}&Oscillations&Oscillations& \textit{Initial Oscillations} &\textit{Initial Oscillations} \\ \hline
\textbf{140 $^{\circ}$C}& \textit{Initial Oscillations}\ &Oscillations& \textit{Initial Oscillations} &\textit{Initial Oscillations}\\ \hline
\end{tabular}
\end{table*}
\begin{table}[b]
\caption{Results of the 300-cc cell simulations at all temperatures. The 300-cc cells had two Rb outlet configurations: (1) the Rb film started at the joint between the outlet and the body (Full Rb Outlet), and (2) the Rb film started at the first bend in the outlet (Partial Rb Outlet). Text in italics again denotes prematurely terminated simulations. Only one temperature parameter was used with the Full Rb Outlet, and it was observed that the Rb began to diffuse back into the cell body. For all temperature parameters, the Partial Rb Outlet configuration appeared to reach a steady-state solution, but a lack of computational resources prevented the complete observation of a steady-state solution. The reported $^{129}Xe$ polarization and laser absorption represent the values modelled at the last time-step of the respective simulation.\label{tab:all300ccresults}}
    \begin{tabular}{c|c|c|}
                                    & \textbf{Full Rb Outlet}   & \textbf{Partial Rb Outlet} \\ \hline
    \textbf{110 $^{\circ}$C}        & \cellcolor{black!25}      & \begin{tabular}{c}Xenon Polarization:6.0\%\\ Laser Absorption: 4.8\%\end{tabular}\\ \hline
    \textbf{120 $^{\circ}$C}        & \cellcolor{black!25}      & \begin{tabular}{c}Xenon Polarization 9.6\%\\Laser Absorption: 7.8\%\end{tabular}\\ \hline
    \textbf{130 $^{\circ}$C}        & \cellcolor{black!25}      & \begin{tabular}{c}Xenon Polarization: 14.4\%\\Laser Absorption: 12.5\%\end{tabular}\\ \hline
    \textbf{140 $^{\circ}$C}        & \textit{Back-Diffusion}   &\begin{tabular}{c}Xenon Polarization:20\%\\Laser Absorption: 19.3\%\end{tabular}\\ \hline
    \end{tabular}
    \label{tab:my_label}
\end{table}

Visualizations of the results from all of the simulations can be found in Ref. \cite{SchrankSupportData2020}. Transient results are stored as movies, and the visualization of HP$^{129}$Xe polarizations from steady-state solutions are stored as images.

As will be developed more fully in Section \ref{sec:discusion}, many of the solutions failed to reproduce experimental behavior. Although this is clearly an indication that the model is not a complete description of the system, the solutions provide some direction as to qualitative reasons for poor cell performance.

\subsection{100-cc Cell Simulation Results}
\begin{figure*}[t]
    \subfloat[\label{fig:100ccresult}]{\includegraphics[width=0.4\textwidth]{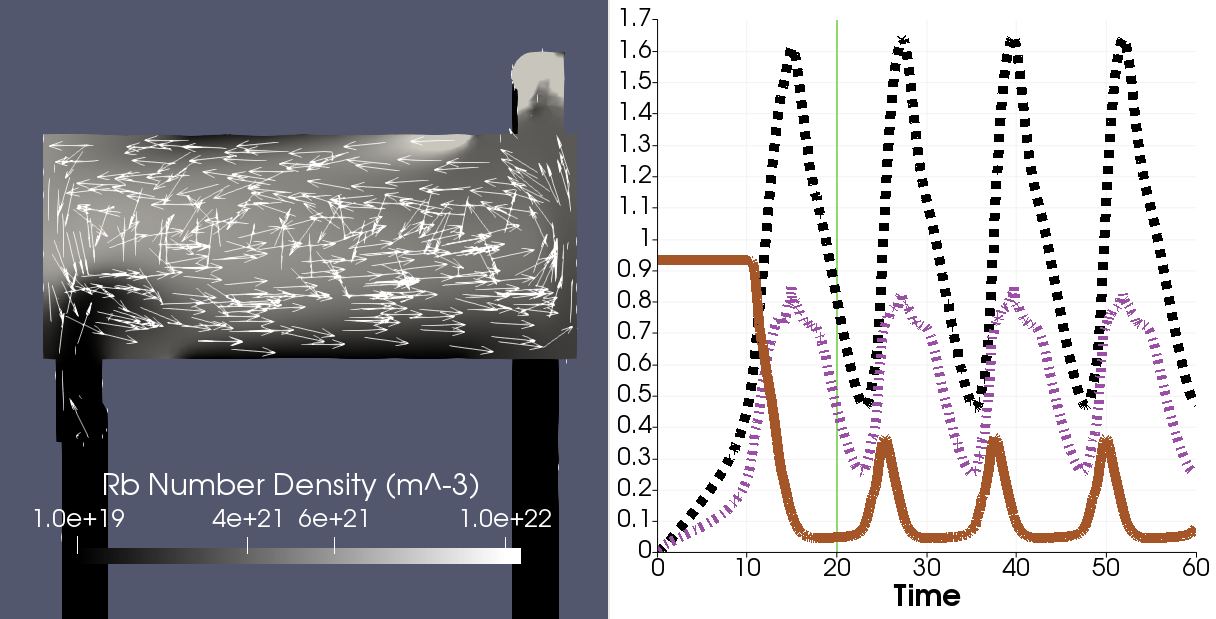}}
    \subfloat[\label{fig:100ccvertresult}]{\includegraphics[width=0.4\textwidth]{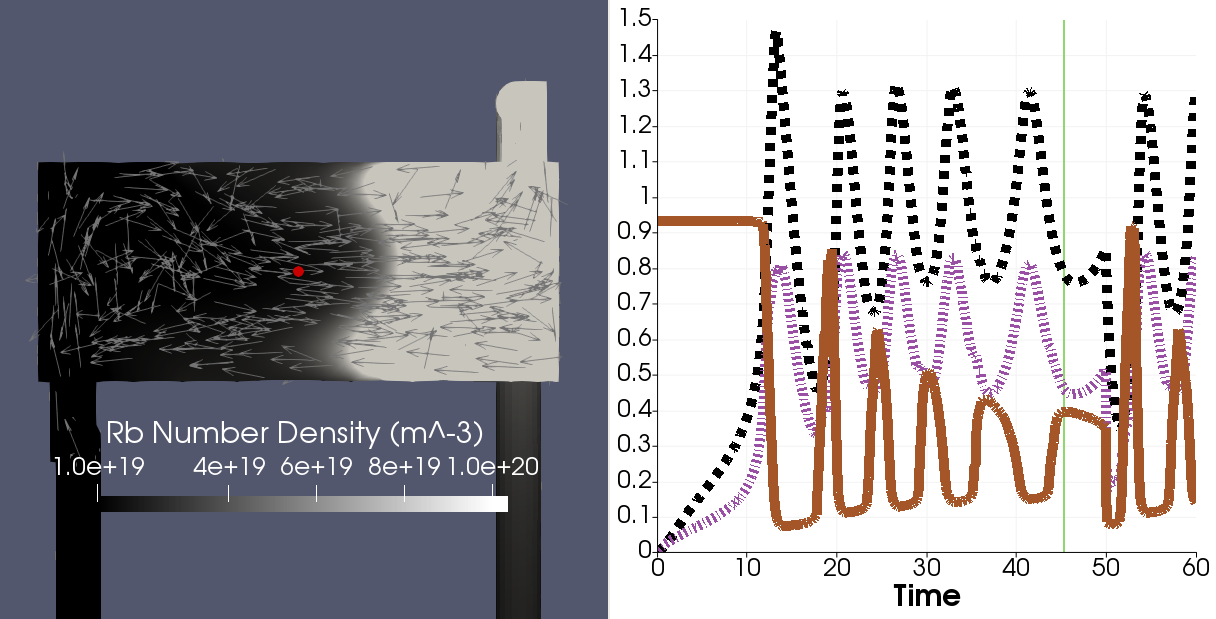}}\\
    \subfloat[\label{fig:300ccrboutletresult}]{\includegraphics[width=0.4\textwidth]{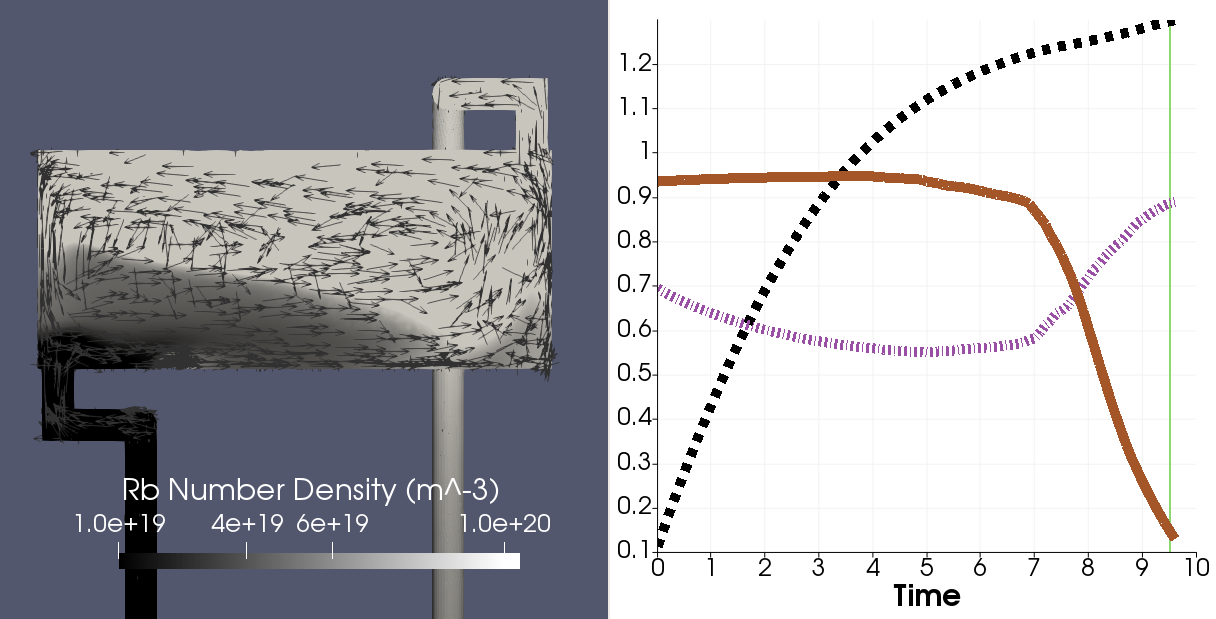}}
    \subfloat[\label{fig:results300ccresult}]{\includegraphics[width=0.4\textwidth]{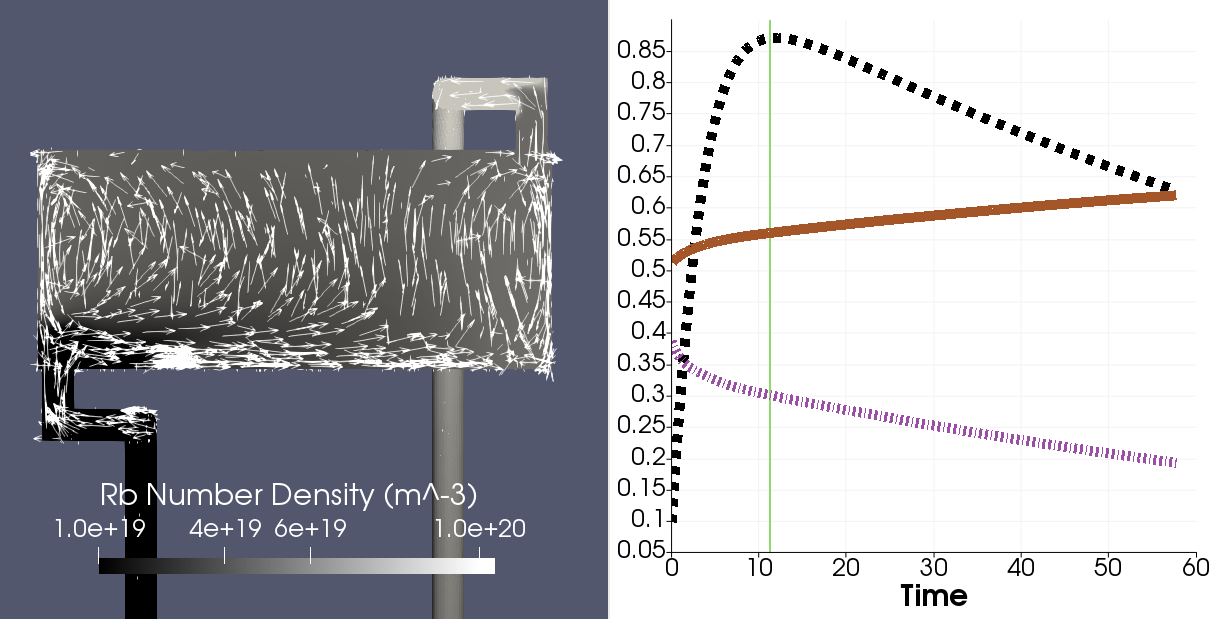}}
    \caption{[Color Online] Four visualizations of the simulations. On the left of each subfigure is a visualization of the optical pumping cell with gas flow streamlines. The velocity vector field of the gas moving through the cells is denoted in each figure. The geometry of the cell is sliced, and the color of the slice denotes the number density of Rb in inverse cubic meters. On the right of each subfigure is a graph of the average temperature in fraction above the set point temperature (black, dashed line), fraction of laser power absorbed (dotted, purple line), and fraction of Rb polarization(solid, brown line). The green vertical line in each graphs denotes the time step at which the visualization on the left is taken.
    The simulations are: (\ref{fig:100ccresult}) the visualization of the 120 $^{\circ}$ C horizontally-oriented 100-cc cell with a full-circumferential Rb distribution; (\ref{fig:100ccvertresult}) the visualization of the 120 $^{\circ}$C vertically-oriented 100-cc cell with a full-circumferential Rb distribution; (\ref{fig:300ccrboutletresult}) the visualization of the initial 140 $^{\circ}$C 300-cc cell with the Rb film in the outlet tube extending up to the wall of the body of the cell; and (\ref{fig:results300ccresult}) the visualization of the 140 $^{\circ}$C 300-cc cell with the Rb film in the outlet tube moved to the first bend of the outlet tube. Solution visualizations for all models and parameters can be found in Ref. \cite{SchrankSupportData2020}.}
    \label{fig:resultvisualization}
\end{figure*}

Summaries of the results for the 100-cc cell simulations are shown in Table \ref{tab:all100ccresults}, and two sample visualizations of the results are shown in Figure \ref{fig:resultvisualization}. 

The 100-cc models in most cases did not reach a steady-state solution. Many of the simulations displayed a strong oscillatory behavior in average temperature, Rb polarization, and laser absorption. Simulations with wall-temperatures above 110 $^\circ$C both displayed this oscillatory behavior and also tended to terminate prematurely, i.e. before the end of the prescribed number of time steps. 

In the case of simulations on horizontally-oriented cells with a full-circumferential Rb source, the trend in oscillatory behavior was definitively seen in all the simulations with a wall temperature above 120 $^{\circ}$C because these simulations did not terminate prematurely. For other cell configurations, the fluid-flow model frequently did not converge at a particular time-step causing the simulations to terminate. In many cases, these premature terminations were restarted only to terminate again due to a similar error. 

When the simulations were successfully restarted, they displayed discontinuities in the calculated laser absorption, average temperature, and Rb polarization. An example of this can be seen in Figure \ref{fig:100ccvertresult} at time step 500 (50 seconds). In this case, all three of the plotted metrics suddenly drop at the time step where the model was restarted. These premature terminations and restart artifacts made it difficult to come to firm conclusions on the long-term behavior of the model for those parameters. 

In some cases, simulations seemed to begin to display oscillatory behavior, but the model terminated prematurely before several oscillations could be observed. However, the suggestion of initial oscillations seemed present. Due to the oscillatory behavior of other simulations of the same cell configuration at lower temperatures, it was assumed, if these simulations could be successfully restarted, they would display the same oscillatory behavior as the simulations at lower temperatures.

Finally, the 120 $^{\circ}$C horizontally-oriented model with a half-circumferential Rb source reached neither steady-state nor displayed oscillatory behavior, but instead displayed a rapid rise in average temperature and laser absorption before terminating prematurely. Again, the premature termination makes it difficult to predict the long-term behavior of the model at these parameters. However, based on the behavior of other simulations at the same temperature, it is suspected that this simulation would have also begun oscillatory behavior as well.

The oscillatory cycles followed a common pattern. During the initial stage, the Rb polarization remained high and stable, but the average cell temperature and laser absorption steadily rose. This state remained until the average gas temperature was roughly 100\% higher than wall boundary condition temperature. At this point, the Rb polarization rapidly decreased, while the laser absorption and temperature rapidly increased.

The average temperature in the cells peaked $\sim$150\% higher than the wall boundary condition, and the Rb polarization fell to $\sim$ 0\%. At this point, the Rb vapor absorbed most of the laser light at the front of the cell. This allowed the temperature of the gas in most of the cell to fall rapidly. As the average temperature decreased, both the Rb vapor number density and laser absorption decreased allowing the laser to, once again, penetrate deep into the optical pumping cell. Thus, the oscillations are driven by the interchange of positive feedback, when laser heated gases vaporize excess Rb, and negative feedback, when the excess Rb blocks laser light at the front of the cell so that it can no longer effectively heat the gas.

The sudden changes described above were accompanied by a marked change in the gas-flow behavior. In the horizontally oriented cells, the convection rotation changed orientations from rotating about the axis of the cell body cylinder to rotating about an axis in the transverse plane (Figure \ref{fig:100ccresult}). This change in flow pattern remained even after the temperature and absorption returned to lower levels. 

Cells oriented vertically show a tightening of the convection rotation towards the front of the cell (Figure \ref{fig:100ccvertresult}). In the case of horizontal cells, this new flow pattern remains even after the average temperature has decreased. In vertical cells, unlike the horizontal models, the change in flow patterns oscillated with the average temperature and laser absorption. 

In addition to this oscillatory behavior, the models predicted points in the cell whose temperature exceed 1000 $^{\circ}$C, which is very likely a non-physical result. If this configuration of Rb is ever realized in an optical pumping cell, the Rb film experiencing high temperatures is likely to relocate to another portion of the cell. In contrast, the Rb sources in the FEM model are inexhaustible and static. 

\subsection{300-cc Cell Simulation Results}
The 300-cc optical pumping cell model was constructed after completing the initial simulations with the 100-cc models. The model's construction attempted to eliminate the above described oscillatory behavior, as it is not observed in experimental setups. First, the volume of the cell was increased in order to decrease the modelled intensity of the laser beam and decrease the likelihood of generating non-physical temperatures. In addition, the Rb source in the the body of the cell was reconfigured to a single drop structure on the bottom surface of the optical pumping cell body. Initially, a Rb sink in the form of a thin film on the walls of the outlet was maintained as in the 100-cc cells. The cell was only simulated in the horizontal orientation. The results are shown in Table \ref{tab:all300ccresults}, and samples of visualization are shown in Figures \ref{fig:300ccrboutletresult} and \ref{fig:results300ccresult}. 

In the initial simulation of this configuration (Figure \ref{fig:300ccrboutletresult}), the model again predicted that the gases in the body of the cell are heated. Due to the configuration and cell dimensions, the temperature near the Rb drop does not significantly rise. However, the gases exiting through the outlet can be several hundred degrees above the set-point temperature.

Excessive heating of the Rb source/sink in the outlet caused an excess of Rb vapor. Due to the proximity of the modelled source to the cell body, some of this excess was able to diffuse back into the cell body and block substantial portions of the light. Eventually, the Rb source in the outlet became the dominant source of Rb vapor in the body of the cell, and the laser adsorption dramatically increased. The simulation terminated prematurely, so it is unclear if this system would generate oscillations as observed in the 100-cc cells.

A second 300-cc geometry was generated with the outlet Rb source/sink recessed farther into the outlet in order to prevent the vapors from this source diffusing back into the body of the cell (Figure \ref{fig:results300ccresult}). The outlet Rb source/sink in this model started at the first bend in the outlet. In this configuration, the hot exit gases again heated the Rb in the outlet and generated excess Rb. However, because the Rb film was recessed farther in the outlet, it did not diffuse into the cell body. At some locations in the outlet, the Rb number density exceed the Rb density in the cell body by an order of magnitude. Due to computational limitations, it was not possible to take these simulations to a steady-state solution.

In the case of the 300-cc geometry, the $^{129}$Xe polarization computational module was enabled. The value for the output $^{129}$Xe polarization and the laser absorption for those simulations is recorded in Table \ref{tab:all300ccresults}. Although the simulations did not meet the definition of steady-state, the trajectory of the observed changes in average cell temperatures, laser absorption, and $^{129}$Xe polarization appeared to be headed towards a steady-state solution. However, limitations in available computational time prevented taking these simulations to more than 600 time-steps.
\section{Discussion of Optical Pumping Cell Simulation Results\label{sec:discusion}}
Because many of the solutions failed to converge to time-independent solutions, the results cannot be used to validate the model against Freeman et. al's results. However, there are insights that can be drawn despite the inability to make a direct comparison with the experimental results.

First, the lack of a steady-state solution is indicative of the dynamics of SEOP flow-through systems. The oscillatory behavior observed in simulated systems is not observed in experimental setups. However, the boundary conditions that seemed to give rise to the oscillatory behavior do not seem entirely unreasonable and are physical possible. Temperatures generated by laser heating in the simulation are high enough to rapidly boil rubidium. That suggests that liquid rubidium sources in the cell body may be dynamically redistributed during SEOP on fast timescales compared to the production of HP$^{129}$Xe. It cannot currently be determined theoretically if there exists a Rb distribution configuration that is stable.

The oscillatory behavior of many of the simulations, while not realized in experimental setups, may also be indicative of the conditions that give rise to laser runaway-heating observed in experimental setups. In experimental SEOP polarizers, laser runaway-heating is characterized by an increase in the absorption of laser-light that gives rise to increased heating of the cell-body \cite{Ruset2005}. It is suspected that the heating creates the same positive-feedback mechanism described above, in which increased heating of the cell-body increases the Rb-vapor density. This increased density absorbs more of the laser light, which in turn heats the cell further.


Second, the simulations indicate that the distribution of Rb in the optical pumping cell body can drastically change the performance of the cell. In particular, Rb sources that are on the ``top'' and ``front'' of the cell are likely to contribute to the Rb vapor concentration to a far greater extent than Rb sources in other areas of the cell. The dominance of Rb sources on the ``top'' was observed in the 100-cc cell simulation in which the Rb source was a full-circumferential distribution (Figure \ref{fig:100ccresult}). This is due to the nature of convection transporting hot gases to the top portion of the cell. Since the gas will be heated by the laser, the system will have a temperature gradient between the upper and lower portions of the cell. The high temperature in the upper portion of the cell will tend to evaporate Rb in those locations more quickly.

Similarly, Rb sources near the ``front'' of the cell are likely to contribute to the concentration of the Rb vapor to a far greater extent than sources in the ``back'' of the cell. Because the laser intensity is highest at the ``front'', gases are likely to be close to the highest temperature experienced in the body of the cell. As seen in the initial 300-cc optical pumping cell simulation, Rb sources that are nominally in the outlet tube may significantly contribute to the Rb number density in the body of the cell (Figure \ref{fig:300ccrboutletresult}).

Depending on the configuration of the Rb in the cell, light may be absorbed preferentially at the front of the cell causing large dark regions in the body. Although not definitively observed in these simulations because the $^{129}$Xe polarization module was disabled, such a situation could give rise to decreased HP$^{129}$Xe polarization.

However, as in the case of runaway-heating, the Rb sources may preferentially relocate to the back of the cell. Therefore, it is unclear the extent of this effect in actual SEOP cells in which the Rb sources are mobile. As cells age, Rb sources may become less mobile and more evenly distributed, and this distribution may contribute to decreased output polarization in aged cells.

Finally, the simulations may indicate that Rb in the outlet of the SEOP cell is of far greater importance than previously suspected. Gases which are significantly warmer than the set-point temperature of the cell may exit the outlet tube and may give rise to Rb number densities in that region that are significantly higher than the Rb number density in the body(Figure \ref{fig:results300ccresult}). Because the laser light does not illuminate the outlet tube of the cell, the Rb vapor in the outlet tube will not be polarized and, thus, may cause rapid depolarization of HP$^{129}$Xe passing through the outlet tube.

It is important to discuss the obvious discrepancy in the observed behavior of the 300-cc cells and reasonable physical behavior of physical SEOP systems, particularly in terms of the laser absorption for a given wall temperature. As described in \cite{Schrank2019ACode}, the model uses a simplified expression for the laser absorption term in the system of differential equations. The absorption presented by the simulated Rb in this configuration may be such that the difference between the simplified expression and the full expression becomes important, which results in a large discrepancy between modelled absorption and those that are observed in physical SEOP systems. 
\section{Conclusion and Future Research\label{sec:conclusion}}
The initial simulations computed using a new FEM model have revealed qualitative insights about current designs for optical pumping cells. In particular, the model indicates that cell performance is strongly linked to the Rb source distribution. Rb sources that are near the ``top'' of the cell or closest to where the pump laser beam enters the cell can contribute more strongly to the overall Rb vapor number density because of heating due to the laser and gas. Even a Rb source that is nominally upstream of the cell body can diffuse back into the body and strongly effect the dynamics of optical pumping. 

Although in this model, the Rb source distribution is static, there is no reason to believe this is the case in actual SEOP systems. Rb metal will redistribute itself into different configurations during the course of the life of the cell, and these configurations may not be optimum for optical pumping. In addition to other mechanisms, this reconfiguration may contribute to the aging and eventual failure of optical pumping cells.

Rb metal films in the outlet tube of the cell may also require careful curation to prevent depolarization in this area. This was first suggested as a problem by Ref. \cite{Ruset2005}, but these simulations suggest the problem may be more acute due to the enhanced number density of Rb vapor in the outlet tube.

All of this taken together suggests that measures should be taken to eliminate Rb metal build up in the outlet tube, or to generate a protocol to occasionally drive the Rb metal from the outlet back into the cell body. This should be considered despite recent innovations to move the Rb source to the inlet tube in a presaturation region, as initially suggested by Ref. \cite{Fink2007}. In fact, the model would suggest that, in the case of optical pumping cells that use a presaturation region, a protocol should be developed to drive Rb from both the cell body and the outlet to preserve control over the Rb number density of the optical pumping region of the cell. 

The subtle dynamics of flow-through optical pumping revealed in these simulations may have application in other areas of spin-exchange physics. For several decades now, Helium-3 polarization has underachieved its theoretical maximum value \cite{Babcock2006LimitsHe3}. The physical mechanism behind the so called X-factor, which describes this deficiency, remains elusive. Perhaps detailed computations of hyperpolarized Helium-3 setups which account for thermal, fluid, and diffusion effects will similarly reveal dynamics that can explain this deficiency.

However, it is clear that additional work needs to be done with the FEM model. The optimal model-solver parameters have not been found such that transient simulations can be run to completion with certainty. This instability, along with the relatively long computational times, makes rigorous computational studies of SEOP cell dynamics challenging. Unfortunately, the nature of FEM makes it difficult to predict optimal solver parameters \textit{a priori}. Thus this problem must be tackled by iteratively attempting different solver parameter combinations.

In addition, because the HP$^{129}$Xe FEM module is unstable in transient simulations, a detailed study of output polarization in simulations that do not reach steady state is not possible. Although experimentally SEOP cells reach steady-state, studying the evolution of the polarization may be useful in the construction of more efficient flow-through polarizer cells.

Finally, since it is clear that the Rb source distribution is critical in the flow dynamics of the cell, it is important to identify physically realistic Rb source distributions. This is important in the FEM simulations as the Rb sources are immobile. Thus, we cannot rely on the computed dynamics of the model to predict changes in the Rb source distribution. 

This FEM model provides insight into understanding the poor performance of current optical pumping systems without the need to appeal to cell contamination or other mechanisms. As improvements are made to the model, it may eventually become necessary to appeal to mechanisms outside of what is described in the model. However, before accepting these explanations, every effort should first be made to understand the behaviour of SEOP systems in terms of known physical principles.
\vspace{0.5cm}
\begin{acknowledgments}
The author gratefully acknowledges J. Cook for useful conversations regarding the coding of the simulations. Also, the author is grateful for insightful discussions with and feedback on initial drafts from B.J. Anderson, T. Barthel, B. Driehuys, and B. Saam. Finally, much of the computation time on Amazon's AWS Cloud Computing Service was supported by backers of the project from Experiments.com. The author gratefully acknowledges the financial support of all the backers of the project.
\end{acknowledgments}

\bibliographystyle{apsrev4-1}

\bibliography{references}
\end{document}